\documentclass[11pt,twoside]{article}

\usepackage{asp2006}
\usepackage{graphicx}
\usepackage{epsf}
\usepackage{psfig}
\usepackage{lscape}

\begin{document}
\title{The VIMOS-VLT Deep Survey: Dependence of Galaxy Clustering on Luminosity}   
\author{A. Pollo\altaffilmark{1,2}, L. Guzzo\altaffilmark{3}, O. Le F\`evre\altaffilmark{1}, B. Meneux\altaffilmark{3,4}, and the VVDS team
}   
\altaffiltext{1}{Laboratoire d'Astrophysique de Marseille, BP8, 13376 Marseille Cedex 12, France} 
\altaffiltext{2}{Astronomical Observatory of the Jagiellonian University, ul Orla 171, 30-244 Krak{\'o}w, Poland}
\altaffiltext{3}{INAF-Osservatorio Astronomico di Brera - Via Brera 28, Milano, Italy}
\altaffiltext{4}{IASF-INAF - via Bassini 15, I-20133, Milano, Italy}

\begin{abstract} 
We have investigated the dependence of galaxy clustering on their intrinsic luminosities at $z \sim 1$, using the data from the First Epoch VIMOS-VLT Deep Survey (VVDS). We have measured the projected two-point correlation function of galaxies, $w_p(r_p)$ for a set of volume-limited samples at an effective redshift $\left<z\right>=0.9$ and median absolute magnitude $-19.6< M_B < -21.3$. We find that the clustering strength is rising around $M_B^*$, apparently with a sharper turn than observed at low redshifts. The slope of the correlation function is observed  to steepen significantly from $\gamma=1.6^{+0.1}_{-0.1}$ to $\gamma=2.4^{+0.4}_{-0.2}$. This is due to a significant change in the shape of $w_p(r_p)$, increasingly deviating from a power-law for the most luminous samples, with a strong upturn at small ($\le 1-2$ $h^{-1}$ Mpc) scales. This trend, not observed locally, also results in a strong scale dependence of the relative bias, $b/b*$ and seems to imply a significant change in the way luminous galaxies trace dark-matter halos at $z\sim 1$ with respect to $z\sim 0$.

\end{abstract}

At the current epoch, luminous galaxies tend to be more clustered than
faint ones, with the difference becoming remarkable above the characteristic 
luminosity $L_*$ of the Schechter luminosity function \citep{norberg02,
zehavi}.  However, so far it was not easy to measure how this effect
evolves with redshift.

To adress this issue, we use data from the first epoch VIMOS-VLT Deep Survey 
\citep[VVDS,][]{LEF04} F02 ``Deep'' field, which is a purely magnitude 
limited survey to $I_{AB}=24$, covering an area of $0.49$ square 
degrees and including $6530$ galaxies with fairly accurate redshifts.   
To measure the dependence of clustering on galaxy luminosity at $z\sim 1$, 
in a broad redshift slice $z \in [0.5,1.2]$ we build a series 
of volume-limited (where statistically possible) or quasi-volume-limited 
sub-samples with typical magnitudes up to $M_B = -21$.

For all these samples we compute the real-space correlation function 
$\xi(r_p,\pi)$ and measure the correlation length $r_0$ and slope 
$\gamma$ from the power-law fit to its projection along the line 
of sight, $w_p(r_p)$, using the method described in \citet{techcorr}. 

We find that both $r_0$ (similarly to what is observed locally) 
and $\gamma$ (unlike observed locally) change with the sample luminosity,
with a strong upturn for samples with luminosities $L>L*$ \citep{lum1}. 
This is produced by a systematic deviation from a simple power-law of 
the shape of $w_p(r_p)$, when looking at more and more luminous samples,
which is seen as a strong upturn on small
($< 1-2$ $h^{-1}$ Mpc) scales, as shown in the left panel of Fig. 1. 

To interpret these results and to compare them to local measurements,  
we compute the relative bias parameter, defined for the generic $i$-th 
sample as $b_i/b^* = \sqrt{(r_0^i)^{\gamma_i}/(r_0^*)^{\gamma^*}}r^{\gamma^*-\gamma^i}$, 
where the $*$ values refer to the $L*$ sample. As shown in the right
panel of Fig. 1, the change of both $r_0$ and $\gamma$ with sample luminosity
results in a strong luminosity dependence of $b/b^*$, which is 
not observed locally. Even more interestingly, this effect depends strongly
also on the particular spatial scale on which it is measured, being more
remarkable at scales smaller than $\sim 4$ $h^{-1}$ Mpc (i.e. "non-linear", according
to a gravitational instability paradigm). This behaviour
indicates a significant change in the way galaxies trace dark-matter halos at $z\sim 1$ with respect to $z\sim 0$ and represents an important constraint for models trying to reproduce small-scale galaxy clustering.

\begin{figure*}[!ht]
\begin{center}
\includegraphics[width = 185pt, height = 185pt]{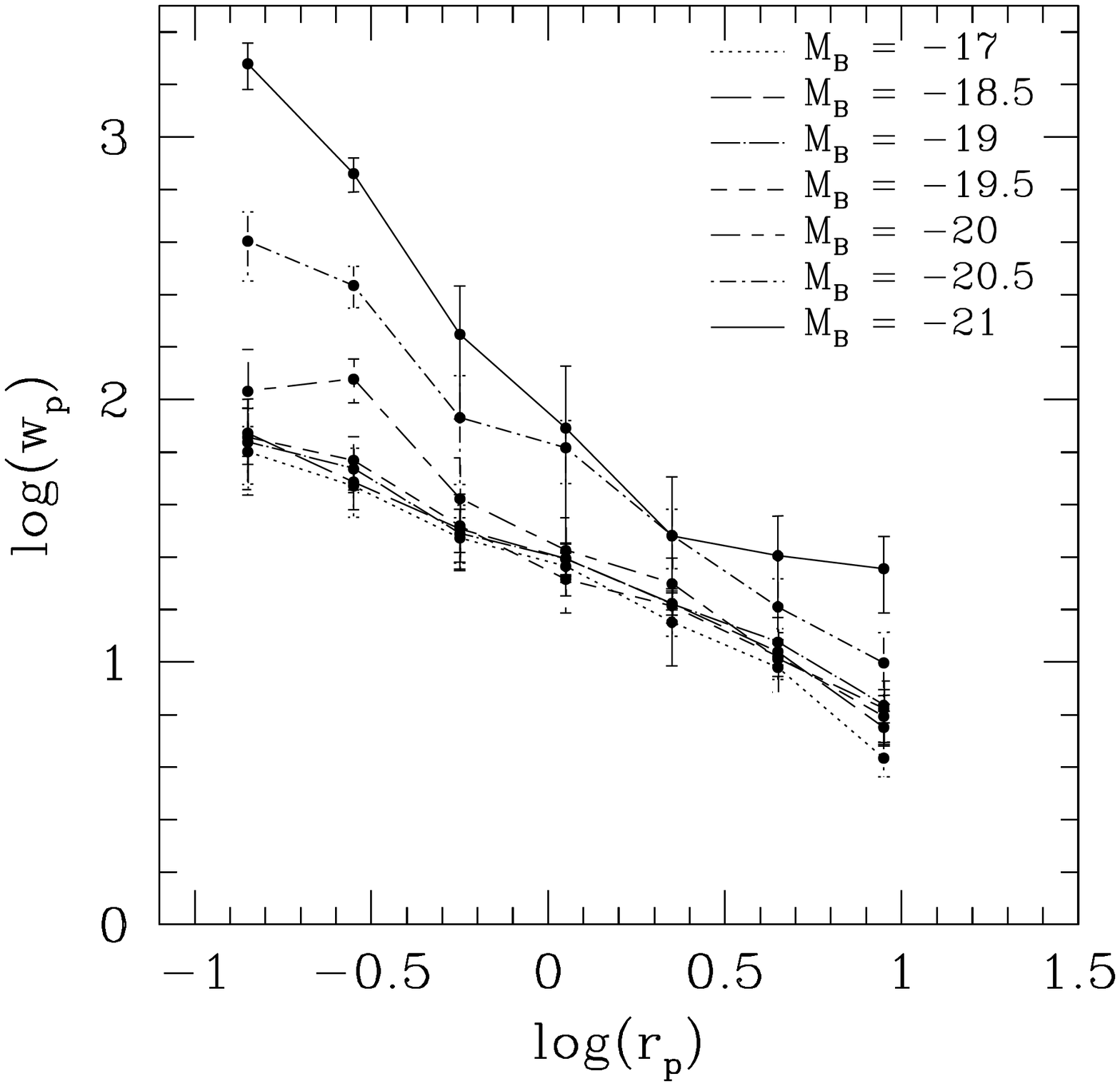}
\includegraphics[width = 185pt, height = 185pt]{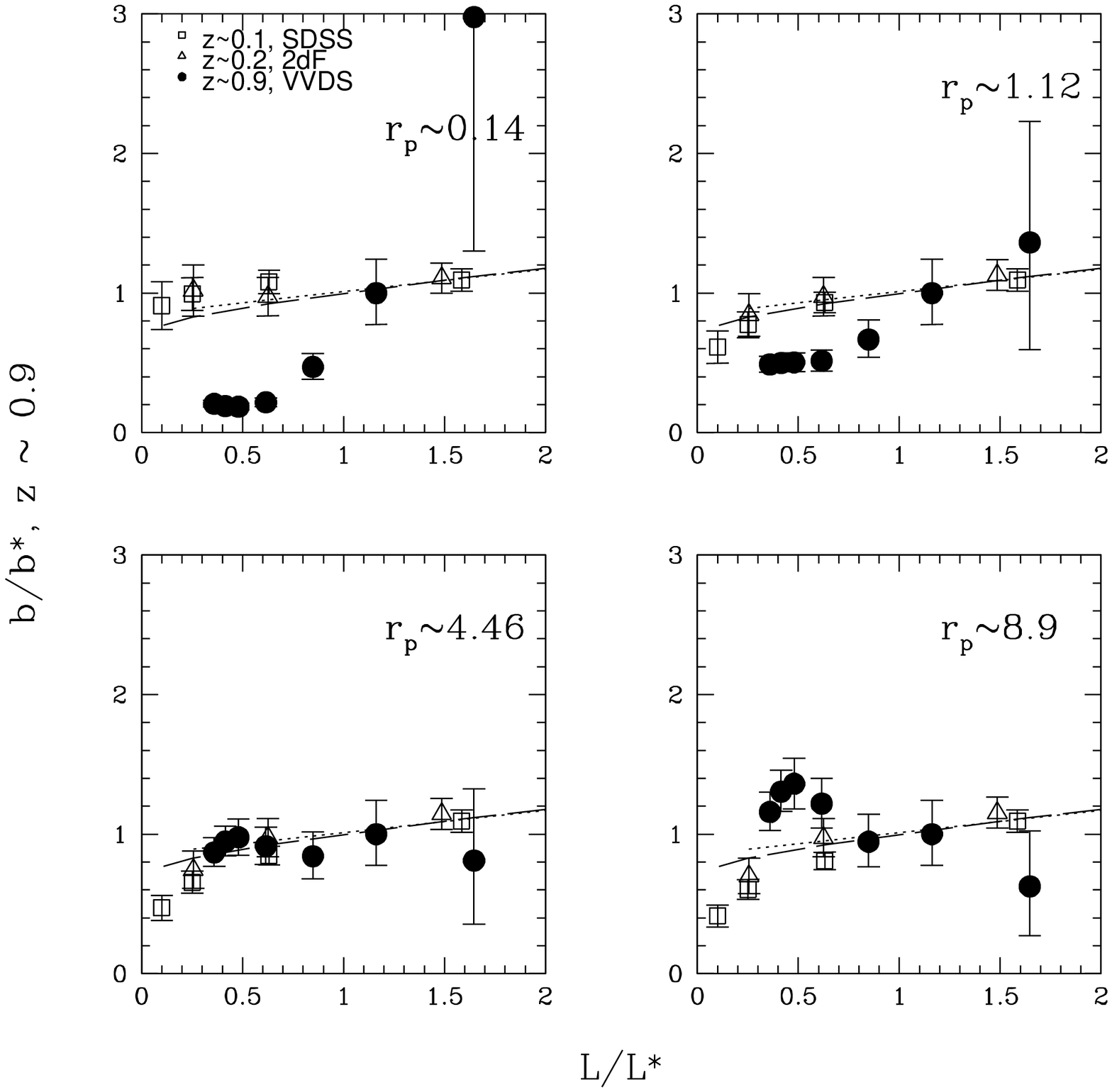}
\end{center}
\caption{Left: Projected 2-point correlation function $w_p$, measured 
at $z \sim 0.9$ for volume-limited samples with limiting absolute 
magnitudes up to $M_B = -21$; its shape for the brightest samples tend 
to deviate from a single power-law fit, with an upturn at small scales. 
Right: As a result, a relative bias strongly depends both on galaxy 
intrinsic luminosities and on a scale on which it is measured, 
unlike observed locally, e.g. in SDSS \citep{zehavi} and 2dF \citep{norberg02}.}\label{fig1}
\end{figure*}


\begin{thebibliography}{}
\bibitem[Le F\`evre et al., 2005]{LEF04} Le F\`evre, O. Vettolani, G., Garilli, B. et al., 2005, A\&A, 439, 845
\bibitem[Norberg et al., 2002]{norberg02}Norberg, P., Baugh, C.M., Haekins, E. et al., 2002, \mnras, 332, 827
\bibitem[Pollo et al., 2005]{techcorr} Pollo, A., Meneux, A. Guzzo, L., et al., 2005, A\&A, 439, 887
\bibitem[Pollo et al., 2006]{lum1} Pollo, A., Guzzo, L., Le F\`evre, O., et al., 2006, A\&A, 451, 409
\bibitem[Zehavi et al., 2005]{zehavi} Zehavi, I., Zheng, Z., Weinberg, D.H. et al., 2005, ApJ, 630, 1
\end{thebibliography}
\end{document}